\documentclass{PoS}

\title{Multi-mission observations of the old nova GK Per during the 2015 outburst}

\ShortTitle{Multi-mission observations of the old nova GK Per during the 2015 outburst}

\author{\speaker{Polina Zemko}\\
        Department of Physics and Astronomy, Universit\'a di Padova, vicolo dell'Osservatorio 3, I-35122 Padova, Italy\\
        E-mail: \email{polina.zemko@studenti.unipd.it}}

\author{Marina Orio\\
        INAF - Osservatorio di Padova, vicolo dell'Osservatorio 5, I-35122 Padova, Italy and Department of Astronomy, University of Wisconsin, 475 N. Charter Str., Madison, WI 53704, USA\\
        E-mail: \email{marina.orio@oapd.inaf.it}}

\author{Gerardo Juan Manuel Luna\\
        Instituto de Astronom\'ia y F\'isica del Espacio, (IAFE, CONICET-UBA), CC 67 - Suc. 28 (C1428ZAA) CABA - Argentina\\
        E-mail: \email{gjmluna@iafe.uba.ar}}

\author{Koji Mukai\\
        CRESST and X-ray Astrophysics Laboratory, NASA Goddard Space Flight Center, Greenbelt, MD 20771, USA and Department of Physics, University of Maryland, Baltimore County, 1000 Hilltop Circle, Baltimore, MD 21250, USA\\
        E-mail: \email{koji.mukai@umbc.edu}}

\abstract{The remarkable old nova and an intermediate polar (IP) -- GK Per was observed with 
 {\sl Swift}, the {\sl Chandra} HETG and {\sl NuSTAR} during its recent dwarf nova (DN)
 outburst in March -- April 2015. Monitoring the outburst,  we noticed 
several processes occurring on different time scales, such as: the slow  
 evolution of the very soft X-ray emission (below 0.6 keV) during the first two weeks of 
the outburst and the very fast saturation of the X-ray flux above 1 keV. The {\sl Swift} UVOT 
lights curves also showed different behaviour, 
depending on the filter. The broad band X-ray spectra revealed the presence of at least 
three different emitting sources. The white dwarf (WD) spin was observed even in the very 
hard X-ray range of {\sl NuSTAR}, indicating that the modulation
is not due to  absorption, in contrast to a typical IP. It is also supported by the 
similarity of the on-pulse and off-pulse X-ray spectra. We propose that the scenario 
when the inner accretion disk pushed towards the WD by the increased accretion obscures the
lower WD pole can work also for GK Per.}

\FullConference{The Golden Age of Cataclysmic Variables and Related Objects - III\\
                 7-12 September 2015\\
                 Palermo, Italy}

\begin{document}

\section{Introduction}
GK Per underwent a nova explosion on 1901 February 21 \cite{wil1901gkper} and after a long period
of irregular fluctuations, in 1948, it started to behave like a DN, with 
a small amplitude (1 -- 3 mag.) outbursts lasting for up to two months, repeated every 
 $\simeq$ 26 months \cite{sab83gkper}.
The most widely accepted explanation of these outbursts is a repeating thermal instability 
in the inner part of the accretion disk (inside-out outbursts; see
\cite{bia86gkper}, \cite{kim92tti} for application of the disk instability to GK Per). GK Per hosts a magnetic
 WD (first proposed by \cite{kin79gkper} and \cite{sab83gkper}), and, hence, the accretion disk is truncated by
the magnetosphere of the WD that surprisingly does not prevent the instability. 
Watson et al. first discovered the X-ray modulation with the period of 351 s, related to the WD spin \cite{wat85gkper}. The authors noticed
that the pulse fraction is remarkably constant --- 50 \% and only at 3 keV is up to 80 \%.

The orbital period is quite long -- 1.997 d \cite{cra86gkperorb} and the distance
to the object is well defined  -- 470 pc \cite{mcl60gkperdist}. The secondary is a K2 type 
subgiant with the mass of 0.25 M$_{\odot}$ (\cite{war76gkper}, \cite{wat85gkper}) and the mass 
of the primary is $\leq$0.72 M$_{\odot}$ \cite{wat85gkper}.

On 2015 March 6.84 Dubovsky (VSNET-ALERT 18388) and Schmeer (VSNET-ALERT 18389) discovered that 
GK Per has started a new DN outburst and was at a magnitude 12.8. We proposed 
a multimission observation campaign in order to follow the evolution of the object  during the 
outburst and to obtain X-ray spectra in a broad energy range, revealing the physical processes 
that take place in this binary system.

\section{Observation and data analysis}
We started the observations of the 2015 DN outburst of GK Per as soon as it became visible
for {\sl Swift} -- on March 12 2015 and observed it almost until optical maximum. 
We obtained two exposures per day with {\sl Swift} for two weeks 
and one exposure per day for another two weeks. Coordinated {\sl NuSTAR} and
{\sl Chandra} Advanced Imaging Spectrometer High-Energy Transmission Grating (ACIS-S/HETG) 
observations were performed on April 4 2015, close to the optical maximum. The {\sl Swift} X-Ray 
Telescope (XRT) and Ultraviolet Optical Telescope (UVOT) data were processed with the 
{\it ftools} package. We used the processed {\sl Swift} Burst Allert Telescope
(BAT) data from the {\sl Swift} BAT transient monitor page \cite{kri13swiftbat}.
We reduced the {\sl Chandra} data with CIAO v.4.7 and the {\sl NuSTAR} data with the standard
{\it nuproducts} pipeline.  All the light curves were extracted after the barycentric 
correction. The X-ray spectra were analysed and fitted using XSPEC v. 12.8.2. The list of the 
observations with the exposure times and count rates is presented in Table 1. 

\begin{table}
 \begin{minipage}{130mm}
\caption{Observational log}
\begin{tabular}{l|ccc}
\hline
\hline
Instrument 			& Date$^a$ 				& Exp.(s)			& Count rate$^b$ \\
\hline	
{\sl Swift} XRT	   		& 57093.15 -- 57121.06  & 373.6 --  2401.5 	& $0.67 - 2.14$ \\
Chandra	MEG		& 57116.83				& 69008  				& 0.0751$\pm$0.0010\\
Chandra	HEG		& 57116.83 				& 69008  				& 0.1214$\pm$0.0013 \\
NuSTAR	FPMA 	& 57116.12 				& 42340  				& 3.665$\pm0.009$  \\
NuSTAR	FPMB 	& 57116.12 				& 42340  				& 3.626$\pm$0.009 \\
\hline	              
\multicolumn{4}{p{.9\textwidth}}{{\bf Notes.}$^a$ Modified Julian Date. $^b$ The count rates
were measured in the following energy ranges: 
 {\sl Chandra} Medium Energy Grating (MEG) - 0.4--5.0 keV, {\sl Chandra} High Energy Gratings (HEG) - 
 0.8--10.0 keV, {\sl NuSTAR} Focal Plane Modules A and B (FPMA and FPMB) - 3--79 keV.}\\
\hline
\hline
\end{tabular} 
\end{minipage}
\end{table}

\section{Results}
The long-term light curves in optical, ultraviolet (UV) and X-rays are presented in fig. 
\ref{fig:lc}. The optical light curve was obtained from the AAVSO\footnote{
American Association of Variable Star Observers: www.aavso.org}. 
The {\sl Swift} XRT count rate was variable by $\sim$50\% but with no significant
increasing or decreasing trend. We probably missed a steep rise of the X-ray count rate, which
is usually observed in the first days of the outburst in GK Per (see \cite{sim15gkper}
for the comparison of the previous eight DN outbursts of GK Per). The high energy
light curve of {\sl Swift} BAT is more stable and only shows a moderate decrease after reaching 
maximum around day 10 after the outburst (AO). It also suggests the outburst may have started
2 days earlier in the {\sl Swift} BAT energy range than in optical. The 
hardness ratio (HR) revealed a gradual softening of the X-ray spectrum, with
minimum around
 20 days AO. The X-ray radiation reaches a plateau already in several days AO, while the optical 
 emission is not saturated at all. The {\sl Swift} UVOT light curves in different filters 
 (the second panel of fig. \ref{fig:lc}) are very different from each other: the 
 U filter light curve is like the hard X-ray light curve, while the UVM2 light curve mimics
the optical one.  

\subsection{Timing analysis}
\begin{figure}
\includegraphics[width=120mm]{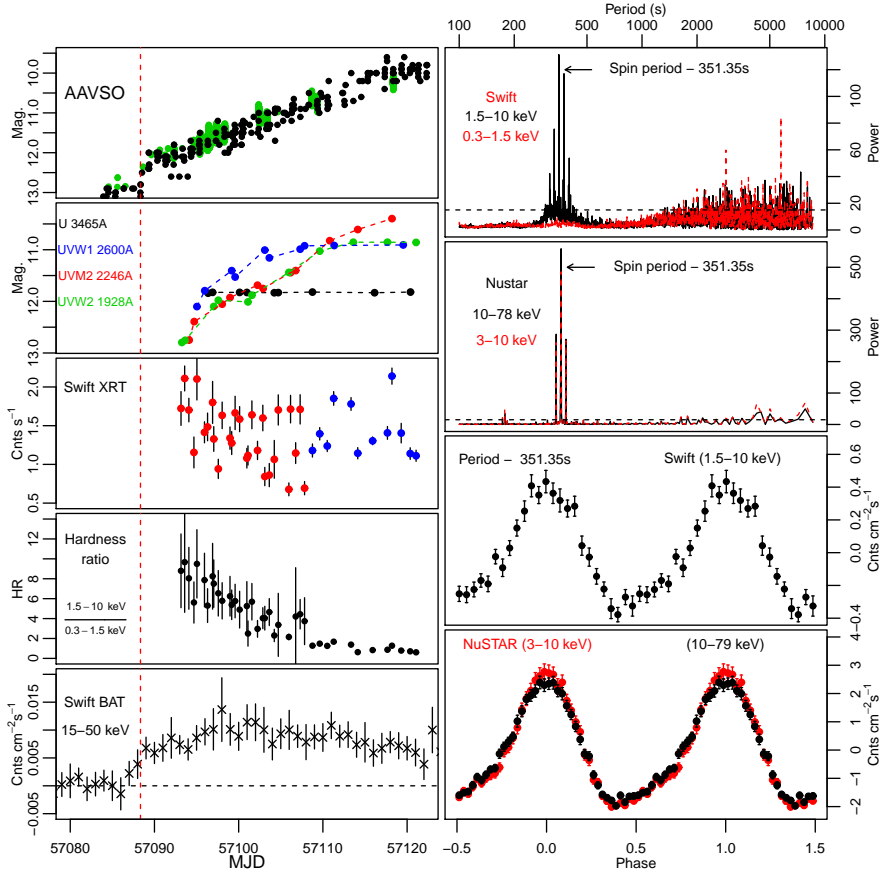}
\caption{Left (from top to bottom): AAVSO light curve in the V band (green) and without 
filter (black). The red vertical line marks the beginning of the outburst in the optical band
in all the panels. The {\sl Swift} UVOT light curves in different filters. The {\sl Swift} XRT light
curve in the photon counting (red) and window timing (blue) modes. The hardness ratio. The {\sl Swift} BAT 
light curve. Right (from top to bottom): The LSPs of the {\sl Swift} XRT light curves below 1.5 keV 
(the red dashed line) and above 1.5 keV (the black line). The LSPs of the {\sl NuSTAR} light curves
below 10 keV (the red dashed line) and above 10 keV (the black line). The {\sl Swift} XRT light 
curve above 1.5 keV folded with the WD spin period. The {\sl NuSTAR} light curves below 10 keV 
(the red points) and above 10 keV (the black points) folded with the WD spin period.}
\label{fig:lc}
\end{figure}

For our timing analysis we extracted the X-ray light curves from the {\sl Swift} XRT, 
{\sl NuSTAR} and {\sl Chandra} observations. In order to investigate a possible energy dependance of a 
variability we subdivided the {\sl Swift} XRT light curves in two energy ranges: 0.3--1.5 keV and
1.5--10 keV and the {\sl NuSTAR} light curves 
were extracted below and above 10 keV. We binned the
{\sl Swift} light curves every 30 s and the {\sl NuSTAR} light curves every 10 s, subtracted the linear
trends and performed the timing analysis using the Lomb-Scargle method \cite{sca82LS}. The Lomb-Scargle periodogramms (LSPs) in a broad
range of periods are presented in the right panels of fig. \ref{fig:lc}. There are very strong
spikes in the {\sl Swift} and {\sl NuSTAR} LSPs, which correspond to the same period -- 351.35 s. We see
from the long term {\sl Swift} XRT light curve that the mean count rate varied from observation 
to observation. Since each exposure covered at least two spin periods this variability is 
not due to the different spin phases. Aiming to remove the contribution from this 
long-term variability, we normalized each individual {\sl Swift} light curve in the harder energy range
to the mean value of 
the count rate within the exposure, combined the light curves and calculated the LSP again. 
The period we found is the same -- 351.35$\pm$0.02. The peak-to-peak pulse amplitude is 
$\sim$ 10 cnts s$^{-1}$ in the {\sl NuSTAR} data and $\sim$ 1 cnts s$^{-1}$ in the {\sl Swift} XRT data,
however, the pulse fraction is the same in both observations -- about 50\%. 

The bottom-right panel of fig. \ref{fig:lc} shows the phase folded {\sl NuSTAR} light curves in two 
energy ranges: 3--10 (red) and 10--79 (black) keV. The modulation is present even in the harder
range and the spin profiles are almost identical. There is a hint that in the hard
X-rays the pulse profile is double-peaked. Typically the spin modulation is not observed in the 
hard X-rays, since the cross section of the photoelectric absorption that causes the modulation 
decreases with energy. A possible mechanism of the pulsations in GK Per will be discussed 
later. 

The LSP of the soft {\sl Swift} XRT light curve (the red dashed line in the top-right panel of 
fig. \ref{fig:lc}) shows a strong spike at 5765 s. Several authors also reported 
 quasi periodic variability on timescales of kiloseconds 
(e.g. \cite{wat85gkper}, \cite{mor99gkperqpo}, \cite{hel04gkper}).

We extracted the light curves from the {\sl Chandra} HETG data in the regions of the strongest 
emission lines of Mg, Si and Fe K$\alpha$ and checked whether
 the flux in these lines is modulated
with the orbital or with the spin period. We found that only the Fe K$\alpha$ 
line emission shows pulsations with the spin period. The other lines do not show any 
measurable periodical modulation, however the flux is quite variable.

\subsection{X-ray spectra}
\begin{figure}
\includegraphics[width=120mm]{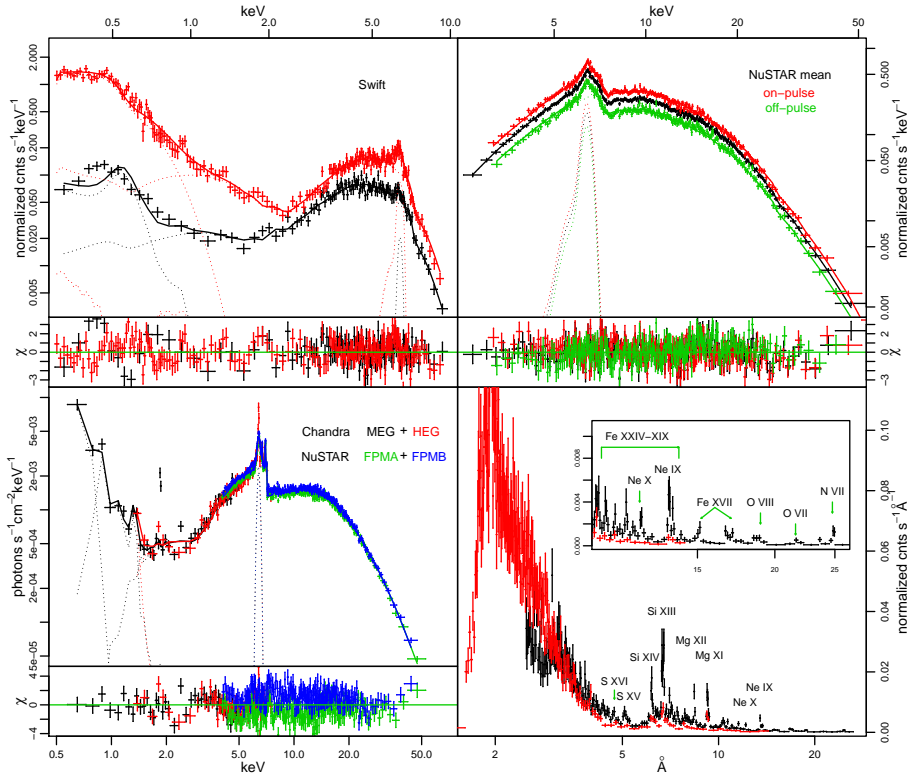}
\caption{Top-left: The {\sl Swift} XRT spectra, integrated over the first two weeks (black) and 
the following two weeks (red) of the observations. Top-right: The {\sl NuSTAR} FPMA mean (black), 
on-pulse (red) and off-pulse spectra. Bottom-left: The combined {\sl Chandra} MEG (black), HEG (red)
and {\sl NuSTAR} FPMA (green) and FPMB (blue) spectra. In all these three panels the solid lines
show a preliminary fit with the following model in XSPEC -- wabs$\times$(bb+pcfabs$\times$pcfabs$\times$(apec+apec+gauss)).
The dashed lines show the model components. Bottom-right: The {\sl Chandra} MEG (black) and HEG (red) 
spectra with the identified emission lines. }    
\label{fig:spec}
\end{figure}

In fig. \ref{fig:spec} we present the X-ray spectra from all the instruments. 
The top-left panel shows the combined {\sl Swift} XRT spectra: the black line is the spectrum 
integrated over the first two weeks of the observations and the red line -- over the 
next two weeks. The main difference between them is in the very soft range, below 0.6 keV, which also  
explains the HR curve (see fig. \ref{fig:lc} ). The combined {\sl Chandra} HETG and {\sl NuSTAR} spectra
are presented in the top-right panel of fig. \ref{fig:spec}. 
This broad band spectrum has three distinct regions, which show different properties. The 
best fitting model that we applied consists of a blackbody and a two-temperature
thermal plasma emission, highly absorbed by the partially covering absorber. The
blackbody component has a temperature of $\sim$25 eV. Although the thermal plasma 
model can marginally represent the shape of the continuum in the 0.8 -- 4 keV range it 
fails to fit the emission lines in the {\sl Chandra} HETG data. The line ratios in the helium-like
triplets of Ne, Mg and Si and the fact that emission from the He-like ions is stronger 
than from the H-like ions indicates that the emitting plasma is not collisional ionized and
there is a possible contribution from the photoionization processes \cite{muk03twotypes}. 
The hard part of the spectrum is well represented by the thermal plasma emission at 
the temperature of 14 keV, but requires a very high value of absorption of the order
of $10^{24}$ cm$^{-2}$. We also extracted the on-pulse and off-pulse spectra from the NuSTAR
observations. The shapes of the spectra were almost the same and only require different 
emission measures of the thermal plasma emission in the best-fitting model. It confirms the
assumption that the spin modulation of the X-rays is not due to absorption.

\section{Discussion}
Mauche analysed all the available estimates of the WD spin period in GK Per and found a possible
decreasing trend, indicating that the WD in GK Per is spinning up \cite{mau04gkper} . We compared our measurements
of the spin period with the predictions of this trend and found that our values are much 
longer. We also re-analysed the {\sl Swift} XRT and UVOT observations of GK Per, obtained during the 
2012 DN outburst and found the same value of spin the period -- 351.35 s, which is not 
consistent with the predicted trend. 

The presence of the spin modulation in the very hard X-rays indicates that it is not due to
absorption in the accretion column, like in a typical IP. Hellier et al. proposed
 that in the eclipsing IP, XY Ari, increased accretion pushes the inner accretion disk inwards,
 blocking the visibility of the secondary accreting pole \cite{hel97XYAri}. As a result, during the DN 
outburst we see only the primary accretion column, and the visibility depends on 
the WD spin phase. This model may also explain our GK Per observations. 

The broad band X-ray spectra revealed the presence of three distinct regions: 
very soft emission at $T_{\rm bb}=25$ eV, absorbed with only interstellar absorption, 
emission in the range 0.8 -- 4 keV, with a possible impact from the photoinization processes
and a highly absorbed thermal plasma emission at T=14 keV. Only the third component of the 
spectrum shows modulation at the spin period. The origin of the very soft component
that gradually develops during the explosion is an open question. 

\section{Acknowledgements}
This work has been partially funded with an ASI-INAF I/037/12/0 award.

\bigskip
\bigskip
\noindent {\bf DISCUSSION}

\bigskip
\noindent {\bf VOITEK SIMON:} Did you detect any changes of the profile and amplitude
of the spin modulation with the progress of the outburst of GK Per?

\bigskip
\noindent {\bf POLINA ZEMKO:} We tried to sum up every 5 observations, extract the 
light curve above 1.5 keV, remove the trend and to fold the resultant light
curves with the spin period. The amplitude of modulation was stable, but the spin
profile became more smooth with time.


\begin{thebibliography}{99}

\bibitem{wil1901gkper}
{Williams}, A.~S. 1901, Astronomische Nachrichten, 155, 29

\bibitem{sab83gkper}
{Sabbadin}, F., \& {Bianchini}, A. 1983, A\&A, 54, 393

\bibitem{bia86gkper}
{Bianchini}, A., {Sabbadin}, F., {Favero}, G.~C., \& {Dalmeri}, I. 1986, A\&A,
  160, 367

\bibitem{kim92tti}
{Kim}, S.-W., {Wheeler}, J.~C., \& {Mineshige}, S. 1992, ApJ, 384, 269

\bibitem{kin79gkper}
{King}, A.~R., {Ricketts}, M.~J., \& {Warwick}, R.~S. 1979, MNRAS, 187, 77P

\bibitem{wat85gkper}
{Watson}, M.~G., {King}, A.~R., \& {Osborne}, J. 1985, MNRAS, 212, 917

\bibitem{cra86gkperorb}
{Crampton}, D., {Fisher}, W.~A., \& {Cowley}, A.~P. 1986, ApJ, 300, 788

\bibitem{mcl60gkperdist}
{McLaughlin}, D.~B. 1960, in Stellar Atmospheres, ed. J.~L. {Greenstein}, 585

\bibitem{war76gkper}
{Warner}, B. 1976, in IAU Symposium, Vol.~73, Structure and Evolution of Close
  Binary Systems, ed. P.~{Eggleton}, S.~{Mitton}, \& J.~{Whelan}, 85

\bibitem{kri13swiftbat}
{Krimm}, H.~A., {Holland}, S.~T., {Corbet}, R.~H.~D., et al. 2013, ApJS, 209, 33

\bibitem{sim15gkper}
{{\v S}imon}, V. 2015, A\&A, 575, A65

\bibitem{sca82LS}
{Scargle}, J.~D., 1982, ApJ, 263, 835

\bibitem{mor99gkperqpo}
{Morales-Rueda}, L., {Still}, M.~D., \& {Roche}, P. 1999, MNRAS, 306, 753

\bibitem{hel04gkper}
{Hellier}, C., {Harmer}, S., \& {Beardmore}, A.~P. 2004, MNRAS, 349, 710

\bibitem{muk03twotypes}
{Mukai}, K., {Kinkhabwala}, A., {Peterson}, J.~R., et al. 2003, ApJ, 586, L77


\bibitem{mau04gkper}
{Mauche}, C.~W. 2004, in Astronomical Society of the Pacific Conference Series,
  Vol. 315, IAU Colloq. 190: Magnetic Cataclysmic Variables, ed. S.~{Vrielmann}
  \& M.~{Cropper}, 120

\bibitem{hel97XYAri}
{Hellier}, C., {Mukai}, K., {Beardmore}, A.~P. 1997, MNRAS, 292, 397


\end{thebibliography}
\end{document}